\documentstyle[12pt]{article}
\newcommand{\be}{\begin{equation}}
\newcommand{\ee}{\end{equation}}
\newcommand{\beq}{\begin{eqnarray}}
\newcommand{\eeq}{\end{eqnarray}}
\begin{document}
\begin{center}

{\bf NON-ABELIAN EIKONAL FORMALISM IN QCD$_4$\footnote{Supported in part by DOE
Grant \# DE-FG02-91ER 40688}}\\
[10mm]

{\bf H. M. FRIED} \\
{\it Physics Department, Brown University}\\
{\it rovidence, RI  02912 USA}\\
\vspace{8mm}
{\bf ABSTRACT}
\end{center}

Using a new technique for the extraction of leading-log energy dependence from
the non-Abelian portion of quark-line interactions, hadronic scattering
amplitudes may be represented in terms of functional integrals over gluonic
material exchanged between scattering quarks.  If that material is further
represented by a condensed tube of gluonic flux (obtained by dimensional
transmutation), these functional integrals will produce effective propagators
corresponding to plasma oscillations, or ``condensons", in this condensed
gluonic material.  These condensons yield a non-zero, non-tachyonic, finite,
gauge-invariant contribution only when the gluonic flux tube corresponds to a
``rigid string" of negligible thickness, so that this formalism provides a
natural mechanism for a ``dynamical string" in QCD4.  Summation over a subset
of  such relevant condensons (at least in SU(2)) generates a form of the
Donnachie-Landshoff Pomeron for the scattering amplitude.  Other,
non-condensate approaches are also available.

\newpage

What can happen when a pair of quarks inside different hadrons have approached
each other sufficiently so that they exchange a gluon?  Immediately, that gluon
splits into two, and then splits again and again into low-frequency gluons
which interact with each other, converting that initial gluon into an entity
which should not be described in terms of individual gluons, but rather in
terms of a semi-classical, gluonic material, or flux.  If that flux forms a
tube with finite energy per unit length, one anticipates that such a flux tube
may, in a non-scattering context, have something to do with confinement.  Note
that the gluonic flux in the tube need have nothing to do with ``classical
QCD", in the sense of satisfying classical field equations.  In fact, these
remarks are valid in a dimensional transmutation (DT) context\cite{one}, as
long as the tubes are understood to be ``rigid", corresponding to flux pointing
in the same spatial direction at every point of the tube.  For technical
reasons, such ``rigidity" is needed in the DT analysis; and it is also
reminiscent of the ``thin needles" proposed a decade ago by
Preparata.\cite{two}

What is the radius of such a tube?  DT (as calculated in the gluonic sector)
does not say; but if the ``plasma oscillation" excitations, or ``condensons" -
which correspond to the relevant degrees of freedom in the tube's condensed
flux - are to be emitted and absorbed by the scattering quarks, then the
condenson propagator can contribute only if the flux tube has a vanishingly
small radius, i.e., if the tube can be thought of as a rigid string.  And the
only portion of the condenson propagator which does contribute is that part
which is non-tachyonic, effectively gauge invariant, and finite.  A summation
over the simplest subset of such relevant condensons can be performed for
SU(2), and produces\cite{three} a scattering amplitude of Donnachie-Landshoff
form.\cite{four}

The above scenario is not the only one possible.  Starting from the non-Abelian
version of the eikonal approximation of quark-line interactions, one can
display $qq$ scattering amplitudes in terms of functional integrations (FIs)
over the gluonic field generated by and exchanged between the scattering
quarks.  In these representations, which contain in principle, the full gluonic
interactions necessary for overall gauge-invariance of the amplitude, the
non-Abelian eikonal analysis\cite{five} allows one to extract, in a
straightforward, functional way - using a new principle called ``contiguity" -
the leading-log energy dependence of the FI integrands, leaving the
momentum-transfer dependence buried in the gluonic fields and hadronic wave
functions.  The challenge is then to extract the proper $s$- and $t$-dependence
of the amplitude, by rescaling the FI variables, by creating a differential
equation in $s$ and $t$ directly from the FI, or by some other, as yet
unappreciated method.  At present, the only results known to the author are
those of the DT analysis described above.

It will perhaps be most efficient to devote the remaining paragraphs of this
note to the new, non-Abelian, quark-line, eikonal representation which
underlies the entire discussion.  The exact scattering amplitude can be cast in
the form of appropriate FIs over Green's functions corresponding to the
scattering quarks, and in the mass-shell-amputated limit, each quark will
produce an ordered-exponential (OE) factor of form

\be
\left( \exp \left[ - ig \int_{-\infty}^{+\infty} \, ds \, p_{\mu} \, A_{\mu}^a
\left( z - sp\right) \lambda_a\right]\right)_+
\ee
which is to be linked, by appropriate FI, to that OE  of the other quark.  The
difficulty is, and has long been, that these factors are OE's, rather than
ordinary exponentials, and the relatively simple structure of Abelian FIs is of
no avail.  Rather, the only obvious, available method for evaluation has been
that of perturbative expansion\cite{six} in the coupling, which is not exactly
what one would prefer to do.  For simplicity, and to restrict the discussion to
its most essential features, let us consider the SU(2) problem where
interactions between the exchanged ``rho-mesons" are neglected, and the only
non-Abelian aspect treated is the coupling of two such OEs, which could be
those of nucleons; here, the coupling is via a FI over a Gaussian (in $A$)
weighting factor.  The FI is trivial in the corresponding Abelian problem, but
impossible to avaluate exactly for our SU(2) example.

The first step is to rewrite each OE in terms of a pair of other FIs (over
proper-time-like variables), with $N'$ a normalization constant,
\be
()_+ = N' \, \int d [\alpha ] \int d[u] \left( e^{i\int_{-\infty}^{+\infty} ds
\lambda_{a} u_{a} (s) } \right)_+ \,\, \cdot \,\, e^{i\int_{-\infty}^{+\infty}
ds \alpha_{a} (s) [u_{a} (s) - gp_{\mu} A_{\mu}^{a} (z-sp)]} \, ,
\ee
so that the $\int d[A]$ FI is Abelian, and may be immediately performed.
Neglecting all self-energy, vertex, $\cdots$ structure along each quark line,
one is then left with the formidable result,
\beq
R & = & N' \int d[\alpha] \int d[u] \, e^{i\int_{-\infty}^{+\infty} ds \,u\cdot
\alpha} \left( e^{i\int_{\-\infty}^{+\infty} ds \lambda^{I} \cdot u}
\right)_+\nonumber \\
& \cdot & N' \int d[\beta] \int d[v] \, e^{i\int_{-\infty}^{+\infty} dt v \cdot
\beta} \,
 \left( e^{i\int_{-\infty}^{+\infty} dt \lambda^{II} \cdot v}\right)_+ \\
& \cdot & \exp \left[ i \int_{-\infty}^{+\infty}\int ds \, dt \, \alpha_a (s)
\, Q_{ab} (s,t) \, \beta_b (t) \right] \, ,\nonumber
\eeq
where in this simplest case $Q_{a,b}$ is the isotopic propagator built out of
the $\rho$-meson propagator,
\begin{displaymath}
Q_{ab} (s\cdot t ) = p_{\mu 1}\, p_{\nu_2} \, \Delta_{c\mu\nu} \left( z_1 -
sp_1 - z_2 + sp_2 \right) ,
\end{displaymath}
and $(z_{\mu 1} , p_{\mu 1})$ and $(z_{\mu 2} , p_{\mu 2})$ are the CM spatial
quark 4-momenta of the scattering problem, in the limit of zero momentum
transfer.

How then does one evaluate or estimate the $R$ of (3)?  For this we have found
it very worthwhile\cite{seven} to ask:  ``What would Schwinger do?" And the
answer is that he would introduce extra parameter and source dependence, in an
effort to find a ``differential characterization" - that is, a differential
equation in parameter and source variables - for $R$.  There are various ways
of doing this, but perhaps the simplest is to replace the $R$ of (3) by the $R
(s;\eta )$ defined as
\beq
R (s ; \eta ) & = & N' \int d[\alpha ] \int d [u ] \,
e^{i\int_{-\infty}^{+\infty} \alpha \cdot u} \left( e^{i\int_{-\infty}^s
\lambda^{I} \cdot u} \right) \cdot \nonumber\\
& \cdot & N' \int d [\beta ] \int d [v] \, e^{i \int_{-\infty}^{+\infty} \beta
\cdot v} \left( s^{i \int_{-\infty}^{+\infty} dt \lambda^{II} \cdot v}\right)_+
\\
& \cdot & e^{i\int_{-\infty}^{+\infty}\int \alpha_{a} Q_{ab} \beta_{b}} \,
\cdot e^{i\int_{-\infty}^{+\infty} dt' v_{b} (t') \eta_{b} (t')} \, \,
,\nonumber
\eeq
where $s$ is a parameter and $\eta(s')$ is a source; clearly, $R = R (s; \eta
)\vert_{s\rightarrow \infty , \eta \rightarrow 0}$.  By calculating $\partial
R/\partial s$, and a sequence of integrations-by-parts, one can obtain a
differential equation linear in $\partial/\partial s$ and $\delta /\delta
\eta$, whose obvious solution is the OE (in $s'$):
\be
R (s; \eta ) = \left( \exp \left[ i \int_{-\infty}^s ds'
\int_{-\infty}^{+\infty} dt' \lambda_a^I \, Q_{ab} (s' , t') \left[ \eta_b (t')
+ \Lambda_b^{II} \left( t' \vert {\delta\over\delta\eta}
\right)\right]\right]\right)_{+(s')} \, \, ,
\ee
\begin{displaymath}
{\rm where} \qquad\qquad\qquad\qquad \Lambda_b^{II} (t\vert iv) = \left(
e^{i\int_{t}^{\infty} \lambda^{II} \cdot v} \right)_+ \, \lambda_b^{II} \left(
e^{-i\int_{t}^{\infty} \lambda^{II} \cdot v}\right) _- \, ,\nonumber
\end{displaymath}
and is an ordered solution of the integral equation
\be
\Lambda_b^{II} \, \left( t \vert {\delta\over\delta\eta} \right) =
\lambda_b^{II} - 2i f_{bcd} \int_t^{\infty} dt' \, {\delta\over\delta\eta_{c}
(t')} \, \Lambda_d^{II} \left( t' \vert {\delta\over\delta\eta} \right) \, .
\ee
Note that the double-ordered $R$ of (3) still contains two sets of enmeshed
orderings; but that they are defined and can be discussed separately.

Solutions to (5) and (6) are not trivial; but there is a new ``contiguity"
technique\cite{five} which extracts that part of $R$ which is clearly going to
be exponentiated (in the form of an ordinary exponential, of course), and which
when expanded in powers of the coupling corresponds to extracting the
leading-log energy dependence of all relevant Feynman graphs.  Further, for our
SU(2) problem, all such terms can be summed without difficulty, and with $R =
\exp [i \chi ]$, generate for the  eikonal $\chi$ of this problem the quantity
\be
\chi = - {g^2\over 2\pi } \left( \sigma^{I} \cdot \sigma^{II} \right) K_0 (\mu
b ) \, \left\{ 1 - {4\over 3} (1 - e^L ) + {2\over 3} L\,\, e^L \right\} \, ,
\ee
\begin{displaymath}
{\rm where}\qquad \qquad L = {g^2\over \pi^2} \, \ln \left( E/m \right) \,\,
K_0 (\mu b ) \,, \qquad {\rm with}\qquad b = z_{\bot 1} - z_{\bot 2}\,
,\nonumber
\end{displaymath}
the impact parameter of the problem.  Here, the Pauli $\sigma$-matrices for
each particle have replaced the Gell-Mann $\lambda_a$ of SU(3).

This eikonal is real, because no (multiperipheral) mechanism for meson emission
has been included; but it does provide a high-energy limit for this limited
SU(2) problem, after a hiatus of more than two decades.

The techniques used above are applicable to other, more modern and relevant
situations, such as the construction of full, QCD scattering amplitudes.  In
particular, the combination of a flux-string description of the gluons with our
$R(s;\eta )$-contiguity technique for the quarks holds out the promise of being
able to begin with standard QCD and to end with realistic scattering
amplitudes.  Of course, the use of non-interacting condensons can only be an
approximation, if a useful one; and an effort should be made to extract those
terms from the full, gluonic stew which will be necessary for the production of
a ``triple Pomeron", as well as Cheng-Wu-type multiperipheral processes
necessary for absolute unitarity restrictions, although the latter could
conceivably be of a non-Froissart form.  One even might expect that an
appropriate, ``static" sum over all non-interacting condensons will be quite
relevant to confinement.  In essence, this is neither phenomenology, nor QCD
lattice magic, but rather an example of analytic, non-perturbative QFT relevant
to experiment.


\begin{thebibliography}{**}
\bibitem{seven}  In other words, one does not have to be a Schwinger in order
to make progress; but, rather, one only need do as he, in other contexts, did.
\end{thebibliography}
\end{document}